\documentclass[aps,showpacs,twocolumn,eqsecnum,superscriptaddress]{revtex4}

\usepackage{amssymb}
\usepackage{amsmath}
\usepackage{latexsym}
\usepackage{graphicx}
\usepackage{longtable}

\newcommand{\gz}{\mbox{\em \r{g}\hspace{0.3mm}}}  

\begin{document}


\title{Multipole expansions for energy and momenta carried by
gravitational waves}

\author{ Milton Ruiz}
\email{ruizm@nucleares.unam.mx}

\affiliation{Instituto de Ciencias Nucleares, Universidad Nacional
Aut\'onoma de M\'exico, A.P. 70-543, M\'exico D.F. 04510, M\'exico.}

\author{Miguel Alcubierre}
\email{malcubi@nucleares.unam.mx}
\affiliation{Instituto de Ciencias Nucleares, Universidad Nacional
Aut\'onoma de M\'exico, A.P. 70-543, M\'exico D.F. 04510, M\'exico.}

\author{Dar\'\i o N\'u\~nez}
\email{nunez@nucleares.unam.mx}
\affiliation{Instituto de Ciencias Nucleares, Universidad Nacional
Aut\'onoma de M\'exico, A.P. 70-543, M\'exico D.F. 04510, M\'exico.}
\affiliation{Current address: Max-Planck-Institut f\"ur Gravitationsphysik,
 Albert Einstein Institut, 14476 Golm, Germany}

\author{Ryoji Takahashi}
\email{ryoji@nucleares.unam.mx}
\affiliation{Instituto de Ciencias Nucleares, Universidad Nacional
Aut\'onoma de M\'exico, A.P. 70-543, M\'exico D.F. 04510, M\'exico.}


\date{\today}


\begin{abstract}
We present expressions for the energy, linear momentum and angular
momentum carried away from an isolated system by gravitational
radiation based on a spin-weighted spherical harmonics decomposition
of the Weyl scalar $\Psi_4$.  We also show that the expressions
derived are equivalent to the common expressions obtained when using a
framework based on perturbations of a Schwarzschild background.  The
main idea is to collect together all the different expressions in a
uniform and consistent way.  The formulae presented here are directly
applicable to the calculation of the radiated energy, linear momentum
and angular momentum starting from the gravitational waveforms which
are typically extracted from numerical simulations.
\end{abstract}



\maketitle


\section{Introduction}
\label{sec:introduction}

Gravitational waves are one of the most important predictions of
General Relativity.  Though such gravitational radiation has not yet
been detected directly, there is strong indirect evidence for its
existence in the form of the now famous binary pulsar PSR~1913+16,
whose change in orbital period over time matches to very high accuracy
the value predicted by General Relativity as a consequence of the
emission of gravitational
waves~\cite{HulseTaylor74,Taylor82a}. Moreover, there is every reason
to believe that the new generation of large interferometric
observatories (LIGO, VIRGO, GEO 600, TAMA) will finally succeed in
detecting gravitational radiation within the next few years.  Also,
gravitational waves are one of the most important physical phenomena
associated with the presence of strong and dynamic gravitational
fields, and as such they are of great interest in numerical
relativity.

Gravitational radiation can carry energy and momentum away from an
isolated system, and it also encodes important information about the
physical properties of the system itself.  The prediction of the
gravitational wave signal coming from the inspiral collision of two
compact objects has been one of the main themes in numerical
relativity over the years, as such predictions can be used as
templates that can significantly improve the possibility of
detection. Nevertheless, it was not until a couple of years ago that
it was finally possible to perform long-term stable numerical
simulations of binary black hole
spacetimes~\cite{Pretorius:2005gq,Campanelli:2005dd,Baker:2005vv,Herrmann:2006ks,Bruegmann:2006at}.
In this way, numerical simulations have now reached a stage where it
is finally possible to extract important astrophysical information
from the collision of two black holes. For instance, it has now become
possible to compute the so-called ``kick'' of the final black hole,
{\em i.e.} the non-zero final velocity of the merged black hole, see
{\em e.g.}~\cite{Bruegmann:2007eb,Pollney:2007ss} and references
therein (we have not attempted to give a complete list of references
here since this field is so active at the moment that such an attempt
would almost certainly become obsolete within a few weeks).  This
non-zero final velocity is a consequence of the fact that in
non-symmetric situations the emitted gravitational waves carry linear
momentum with them. Similarly, the spin of the final black hole can
also be computed by subtracting the angular momentum carried away by
gravitational waves from the initial ADM angular momentum of the
spacetime.

Historically, there have been two main approaches to the extraction of
gravitational wave information from a numerical simulation. For a
number of years the traditional approach has been based on the theory
of perturbations of a Schwarzschild spacetime developed originally by
Regge and Wheeler~\cite{Regge57}, Zerilli~\cite{Zerilli70}, and a
number of other authors, and later recast as a gauge invariant
framework by Moncrief~\cite{Moncrief74}. In recent years, however, it
has become increasingly common in numerical relativity to describe
gravitational waves in terms of the components of the Weyl curvature
tensor with respect to a frame of null vectors~\cite{Campanelli99},
using what is known as the Newman--Penrose formalism~\cite{Newman62a}.
In this paper we obtain in a simple way general expressions for the
energy, linear momentum and angular momentum carried away by
gravitational waves using the spin-weighted spherical harmonics
decomposition of the components of the Weyl curvature tensor. We also
compare these expressions with the more common expressions used in
numerical relativity which are based on the theory of gauge invariant
perturbations of Schwarzschild~\cite{Thorne80b} (see {\em e.g.}
~\cite{Pollney:2007ss,Sopuerta:2006wj}). Though most of the main ideas
and results presented here are known, our aim is to collect all
relevant expressions together using a consistent set of conventions
and definitions.

This paper is organized as follows.  In Sec.~\ref{sec:GW_physics} we
summarize the general expression for the energy and momenta of
gravitational waves in the transverse-traceless (TT) gauge. In
Sec.~\ref{sec:weyl} we obtain the general expressions for the energy
and momenta using the Weyl scalars and the spin-weighted spherical
harmonics.  Later, in Sec.~\ref{sec:GW_perturbations} we briefly
discuss some of the ideas behind the theory of perturbations of a
Schwarszchild black hole and we obtain again the general expression
for energy and momenta in terms of the gauge invariant ``master''
functions $(\Psi_{\rm even},\Psi_{\rm odd})$ and $(Q_{\rm even},Q_{\rm
odd})$. We conclude in Section~\ref{sec:conclusions}.  In addition, in
Appendix~\ref{app:harmonics} we discuss some important properties of
the spin-weighted spherical harmonics.


\section{Gravitational waves and the physics in them encoded}
\label{sec:GW_physics}

Let us start by considering a small perturbation $h_{\mu \nu}$ to a flat
background metric $\gz_{\mu\nu}$, such that the full metric is given
by
\begin{equation}  
g_{\mu\nu} = \gz_{\mu\nu} + h_{\mu\nu} \; ,
\label{eq:pertubation}
\end{equation}
where $|h_{\mu \nu}| \ll 1$.  Knowing that the gravitational field has
only two degrees of freedom, we can choose a gauge such that there are
only two unknown functions, $h^+$ and $h^\times$, that represent the
two possible polarizations of the gravitational waves. That is, we
work in the so-called {\em transverse-traceless} (TT) gauge in which
imposes the conditions $h_{0\alpha}={h^i}_i={h^{ij}}_{\mid j}=0$
(where the bar denotes covariant derivative with respect to the
background metric). The perturbation tensor then has the form
\begin{equation}
{h^{TT}}_{\mu\nu}=\left( \begin{array}{cccc}
0 & 0 & 0 & 0\\
0 & h^+ & h^\times & 0\\
0 & h^\times & -h^+ & 0\\
0 & 0 & 0 & 0
\end{array} \right) \; ,
\label{eq:htt}
\end{equation}
where we have assumed that we have a plane wave propagating along the
$z$ axis.  Moreover, we will be considering only outgoing waves which
means that the functional dependence on $r$ and $t$ of the metric
perturbations is of the form $f(r-t)/r$, so that asymptotically one
has \mbox{$\partial_r h \sim - \partial_t h \equiv - \dot{h}$}.
Furthermore, for large $r$ the radiation can always be locally
approximated as a plane wave, so that angular derivatives can be
neglected when compared with radial derivatives (but one must be
careful when one deals with quantities that do not involve radial
derivatives such as the angular momentum, see below).

Having described the properties of the metric perturbation $h_{\mu
\nu}$, we proceed to find the flux of energy and momentum carried away
by the gravitational waves. The most straightforward way is to use the
Isaacson stress-energy tensor (for details see {\em
e.g.}~\cite{Misner73}), which describes the energy and momentum
associated with the gravitational waves averaged over a few
wavelengths using the so-called short wavelength approximation (one
must remember that in general relativity there is no local expression
for the energy and momentum of the gravitational field). In the TT
gauge, and in a locally inertial frame, the Isaacson stress-energy
tensor is given by
\begin{equation}
T_{\mu \nu} = \frac{1}{32\,\pi} \left< \partial_{\mu} h^{TT}_{ij} \:
\partial_{\nu} h^{TT}_{ij} \right> \; ,
\label{eq:stressenergy}
\end{equation}
where $\left< \: \right>$ denotes an average over several wavelengths,
and where a summation over the repeated indices $(i,j)$ is implied.
Using the explicit form of the $h^{TT}_{ij}$ in terms of $h_+$ and
$h_\times$, the Isaacson stress-energy tensor can be rewritten as
\begin{equation}
T_{\mu \nu} = \frac{1}{16\,\pi}
\left< \partial_\mu h^+ \partial_\nu h^+
+ \partial_\mu h^\times \partial_\nu h^\times \right> \; ,
\label{eq:tTT_1}
\end{equation}
or equivalently (here and in what follows $\bar{z}$ will denote the
complex conjugate of $z$)
\begin{equation}
T_{\mu \nu} = \frac{1}{16\,\pi} \: {\rm Re}
\left< \partial_\mu H \partial_\nu \bar{H} \right> \; ,
\label{eq:tTT_2}
\end{equation}
with $H := h^+ - i h^\times$, and where ${\rm Re}(z)$ denotes the real
part of $z$.

We can now use equation~\eqref{eq:tTT_2} to find the flux of energy
and momentum carried away by the gravitational waves. Consider first
the flux of energy along the direction $i$, which is given in general
by $T^{0i}$. In particular, the energy flux of the gravitational waves
along the radial direction will then be given in locally Cartesian
coordinates by
\begin{eqnarray}
\frac{dE}{dt\,dA} &=& T^{0r} = \frac{1}{16\, \pi} \: {\rm Re}
\left< \partial^0 H \partial^r \bar{H} \right> \nonumber \\
&=& - \frac{1}{16\, \pi} \: {\rm Re}
\left< \partial_t H \partial_r \bar{H} \right> \; ,
\end{eqnarray}
with $dA$ the area element normal to the radial direction.  Using
now the relation $\partial_r h = - \dot{h}$ mentioned
above, we can rewrite this as
\begin{equation}
\frac{dE}{dt\,dA} = \frac{1}{16\, \pi} \: \left< \dot{H} \dot{\bar{H}} \right>
= \frac{1}{16\, \pi} \: \left< | \dot{H} |^2 \right> \; .
\end{equation}
If we want the total flux of energy leaving the system at a given time
we need to integrate over the sphere to find
\begin{equation}
\frac{dE}{dt} = \lim_{r \rightarrow \infty} \frac{r^2}{16\,\pi} \oint
| \dot{H} |^2 d \Omega \; ,
\label{eq:ETT}
\end{equation}
where we have taken $dA = r^2\,d \Omega$, with $d \Omega$ the standard
solid angle element, and where the limit of infinite radius has been
introduced since the Isaacson stress-energy tensor is only valid in
the weak field approximation.  Notice also that we have dropped the
averaging since the integral over the sphere is already performing an
average over space, plus the expression above is usually integrated
over time to find the total energy radiated which again eliminates the
need to take a time average.

Consider next the flux of linear momentum which corresponds to the
spatial components of the stress-energy tensor $T^{ij}$.  The flux of
momentum $i$ along the radial direction will then be given by
\begin{eqnarray}
\frac{dP_i}{dt\,dA} &=& T_{ir} = \frac{1}{16\, \pi} \: {\rm Re}
\left< \partial_i H \partial_r \bar{H} \right> \nonumber \\
&=& \frac{1}{16\, \pi} \: l_i \left< | \dot{H} |^2 \right> \; ,
\end{eqnarray}
where in the last inequality we have used the fact that asymptotically
\mbox{$\partial_i H \simeq (x_i/r) \: \partial_r H$} ({\em i.e.} we
are ignoring angular derivatives), and also the relation between the
radial and temporal derivatives for outgoing waves.  The vector
$\vec{l}$ introduced above is the unit radial vector in flat space,
\begin{equation}
\vec{l} = \vec{x}/r = (\sin \theta \cos \varphi, \sin \theta \sin \varphi,
\cos \theta) \; .
\label{eq:li}
\end{equation}
The total flux of momentum leaving the system will again be given by
an integral over the sphere as
\begin{equation}
\frac{dP_i}{dt} = \lim_{r \rightarrow \infty} \frac{r^2}{16 \pi} \oint
l_i \: | \dot{H} |^2\, d \Omega \; .
\label{eq:PTT}
\end{equation}

Finally, we consider the flux of angular momentum.  Locally, the flux
of the $i$ component of the angular momentum along the radial
direction should correspond to $\epsilon_{ijk} x^j T^{kr}$ with
$\epsilon_{ijk}$ the three-dimensional Levi--Civita antisymmetric
tensor (this is just $\vec{r} \times \vec{p}$ in three-dimensional
notation). However, in the case of gravitational waves this expression
is in fact wrong since the averaging procedure that is used to derive
the Isaacson stress-energy tensor ignores terms that go as $1/r^3$,
and it is precisely such terms the ones that contribute to the flux of
angular momentum.  A correct expression for the flux of angular
momentum due to gravitational waves was first derived by DeWitt in
1971, and in the TT gauge has the form (see {\em
e.g.}~\cite{Thorne80b})
\begin{equation}
\frac{d J^i}{dt \, dA} = \frac{1}{32\,\pi} \: \epsilon^{ijk} \Big( x_j \,
\partial_k h_{ab} + 2 \, \delta_{aj} \, h_{bk} \Big) \, \partial_r h^{ab} \; .
\label{eq:angularmomentumflux_GW0}
\end{equation}
The last expression can be rewritten in more compact form if we
introduce the angular vectors $\vec{\xi}_i$ associated to rotations
around the three coordinate axis.  These vectors are Killing fields of
the flat metric, and in Cartesian coordinates have components given by
$\xi^k_i = {\epsilon_i}^{jk} x_j$ (where $\xi^k_i$ represents the $k$
component of the vector $\vec{\xi}_i$). In terms of the vectors
$\vec{\xi}_i$ the flux of angular momentum can now be written as
\begin{equation}
\frac{d J_i}{dt} = - \lim_{r \rightarrow \infty} \frac{r^2}{32 \pi} \oint
\left( \pounds_{\xi_i} h_{ab} \right) \: \partial_t h^{ab} \: d \Omega \; ,
\label{eq:angularmomentumflux_GW1}
\end{equation}
where $\pounds_{\xi_i} h_{ab}$ is the Lie derivative of $h_{ab}$ with
respect to $\vec{\xi}_i$, and where we have again taken
\mbox{$\partial_r h = - \partial_t h$} for outgoing waves.  The
appearance of the Lie derivative is to be expected on physical
grounds, since for rotational symmetry around a given axis
($\pounds_{\xi_i} h_{ab} =0$) one should find that the corresponding
angular momentum flux vanishes.

In order to write the angular momentum flux in terms of $H$ as we have
done with the flux of energy and linear momentum, one must now
carefully consider the action of the Lie derivative on the perturbed
metric.  The easiest way to do this is to work in spherical
coordinates $(r,\theta,\varphi)$, in which case the angular vectors
have components
\begin{eqnarray}
\vec{\xi}_x &=& \left( 0, - \sin \varphi,
- \cos \varphi \cot \theta \right) \; , \\
\vec{\xi}_y &=& \left( 0,\cos \varphi,
- \sin \varphi \cot \theta \right) \; , \\
\vec{\xi}_z &=& \left( 0, 0, 1 \right) \; .
\end{eqnarray}
It is clear that the vector $\vec{\xi}_z$ corresponds to one of the
vectors of the coordinate basis, which implies that Lie derivatives
along it reduce to simple partial derivatives. To calculate the Lie
derivatives in the other directions it is convenient to first
introduce the two complex angular vectors $\vec{\xi}_\pm :=
\vec{\xi}_x \pm i\, \vec{\xi}_y$. Furthermore, we will introduce an
orthonormal spherical basis
$(\hat{e}_r,\hat{e}_\theta,\hat{e}_\varphi)$, and define the two unit
complex vectors $\hat{e}_\pm := {1}/{\sqrt{2}} \left( \hat{e}_\theta
\mp i\, \hat{e}_\varphi \right)$.  One can then show after some
algebra that the Lie derivative of $\hat{e}_\pm$ with respect to
$\vec{\xi}_\pm$ is given by
\begin{equation}
\pounds_{\xi_\pm} e^a_\pm = \mp \left( i\,e^{\pm i \varphi} \csc \theta
\right) e^a_\pm \; .
\end{equation}

Let us now rewrite the metric perturbation $h_{ab}$ in the TT gauge in
terms of the orthonormal basis as:
\begin{eqnarray}
h_{ab} &=& h^+\, \Big[ (\hat{e}_\theta)_a (\hat{e}_\theta)_b 
- (\hat{e}_\varphi)_a (\hat{e}_\varphi)_b  \Big] \nonumber \\
&+& h^\times\, \Big[ (\hat{e}_\theta)_a (\hat{e}_\varphi)_b
+ (\hat{e}_\varphi)_a (\hat{e}_\theta)_b \Big] \nonumber \\
&=& H \: (\hat{e}_-)_a (\hat{e}_-)_b
+ \bar{H} \: (\hat{e}_+)_a (\hat{e}_+)_b \; .
\end{eqnarray}

We are now in a position to calculate the Lie derivative of $h_{ab}$
with respect to $\vec{\xi}_\pm$. One finds,
\begin{equation}
\pounds_{\xi_\pm} h_{ab} = (\hat{e}_-)_a (\hat{e}_-)_b \: \hat{J}_\pm H
+ (\hat{e}_+)_a (\hat{e}_+)_b \: \hat{J}_\pm \bar{H} \; ,
\end{equation}
where we have defined the operators
\begin{eqnarray}
\hat{J}_\pm &:=& \xi^a_\pm \partial_a - i\,s \: e^{\pm i \varphi} \csc \theta
\nonumber \\
&=& e^{\pm i \varphi}\,\Big[ \pm i\,\partial_\theta - \cot \theta \:
\partial_\varphi - i\,s \csc \theta \Big] \; ,
\end{eqnarray}
with $s$ the spin weight of the function on which the operator is
acting: $s=-2$ for $H$, and $s=+2$ for $\bar{H}$ (see
Appendix~\ref{app:harmonics}). The last result implies that
\begin{eqnarray}
(\pounds_{\xi_\pm} h_{ab} ) \: \partial_t h^{ab}
&=& \hat{J}_\pm H \: \partial_t \bar{H}
+ \hat{J}_\pm \bar{H} \: \partial_t H \nonumber \\
&=& 2\,{\rm Re} \left\{ \hat{J}_\pm H \: \partial_t \bar{H} \right\} \; ,
\end{eqnarray}
from which we find
\begin{eqnarray}
(\pounds_{\xi_x} h_{ab} ) \: \partial_t h^{ab} &=&
2\,{\rm Re} \left\{ \hat{J}_x H \: \partial_t \bar{H} \right\} \; , \\
(\pounds_{\xi_y} h_{ab} ) \: \partial_t h^{ab} &=&
2\,{\rm Re} \left\{ \hat{J}_y H \: \partial_t \bar{H} \right\} \; .
\end{eqnarray}
Collecting results, the flux of angular momentum becomes
\begin{equation}
\frac{d J_i}{dt} = - \lim_{r \rightarrow \infty} \frac{r^2}{16 \pi}
\: {\rm Re} \oint \hat{J}_i H \: \partial_t \bar{H} \: d \Omega \; ,
\label{eq:angularmomentumflux_GW2}
\end{equation}
with the angular momentum operators $\hat{J}_i$ defined as
\begin{eqnarray}
\hat{J}_x &=& \frac{1}{2}\,\left( \hat{J}_+ + \hat{J}_- \right) \nonumber \\ 
&=& - \sin \varphi \: \partial_\theta - \cos \varphi \left( \cot \theta \:
\partial_\varphi + i\,s \csc \theta \right)  \: , \nonumber \\
\hat{J}_y &=& -\frac{i}{2} \left( \hat{J}_+ - \hat{J}_- \right) \nonumber \\
&=& + \cos \varphi \: \partial_\theta - \sin \varphi \left( \cot \theta \:
\partial_\varphi + i\,s \csc \theta \right) \: , \nonumber \\
\hat{J}_z &=& \partial_\varphi \; .\label{eq:Ji}
\end{eqnarray}
Notice that, except for a factor of $-i \hbar$, these are just the
quantum mechanical angular momentum operators with the correct spin
weight~\cite{Dray85}.

We now have expressions for the radiated energy, linear momentum, and
angular momentum~(more details on the derivation of these expressions
can be found in~\cite{Alcubierre07a}).  However, it turns out that the
TT coefficients $h^+$ and $h^\times$ are not trivial to obtain from a
numerical simulation. One of the reasons for this is that during a
numerical simulation one obtains the full spacetime metric, and not
the background spacetime plus a separate perturbation. One therefore
needs to relate these metric perturbations to some geometric
quantities, preferably scalars, that can be obtained directly from the
available data.  In the following Section, we will concentrate on one
such quantity that turns out to be ideal for extracting gravitational
wave information, namely the Weyl scalar $\Psi_4$.


\section{The Weyl scalar $\Psi_4$}
\label{sec:weyl}

\subsection{Definition of $\Psi_4$}
\label{sec:psi4}

It is well known that the scalars needed to completely characterize a
given spacetime can be expressed as projections onto a null tetrad of
the Weyl tensor. These scalars, known as the {\em Weyl scalars}, have
several interesting properties. For instance, for a suitable chosen
tetrad, they can be related directly with the gravitational waves at
null infinity. We will be interested mainly in the Weyl scalar
$\Psi_4$, which is associated with outgoing gravitational radiation
and is defined as
\begin{equation}
\Psi_4 := C_{\alpha \beta \mu \nu} \: k^\alpha \bar{m}^\beta k^\mu
\bar{m}^\nu \; ,
\end{equation}
with $C_{\alpha \beta \mu \nu}$ the Weyl tensor
and where ${k}^\mu$ and ${\bar{m}^\mu}$ are two vectors of a null
tetrad constructed from the orthonormal spherical basis in the
following way
\begin{eqnarray}
l^\mu &:=& \frac{1}{\sqrt{2}} \left( \hat{e}_{t}^\mu
+ \hat{e}_{r}^\mu \right) \; , \nonumber \\
k^\mu &:=& \frac{1}{\sqrt{2}} \left( \hat{e}_{t}^\mu
- \hat{e}_{r}^\mu \right) \; , \nonumber \\
m^\mu &:=& \frac{1}{\sqrt{2}} \left( \hat{e}^\mu_{\theta}
+ i \hat{e}^\mu_{\varphi} \right) \; ,\nonumber \\
\bar{m}^\mu &:=& \frac{1}{\sqrt{2}} \left( \hat{e}^\mu_{\theta}
- i \hat{e}^\mu_{\varphi} \right) \;,
\label{eq:nullt}
\end{eqnarray}
where the vectors $\hat{e}_t^\mu$, $\hat{e}_{r}^\mu$, 
$\hat{e}_{\theta}^\mu$, and $\hat{e}^\mu_{\varphi}$ 
are the usual orthonormal basis induced by the spherical coordinates.
The Weyl scalars $(\Psi_0,\Psi_1,\Psi_2,\Psi_3)$ are similarly defined
as different contractions of the Weyl tensor with the null tetrad.

It turns out that the complex quantity \mbox{$H= h^+ - i h^\times$},
defined in the previous Section, can in fact also be written in terms
of the Weyl scalar $\Psi_4$.  In order to see this notice first that
if we are in vacuum far from the source of the gravitational waves the
Weyl and Riemann tensors coincide (the Ricci tensor vanishes). Using
now the standard expression for the Riemann tensor in the linearized
approximation one can easily show that, for outgoing plane waves in the
TT gauge traveling along the $r$ direction, $\Psi_4$ takes the simple
form
\begin{equation}
\Psi_4 = - \left(\ddot{h}^+ - i\, \ddot{h}^\times\right) = - \ddot{H} \; ,
\label{eq:H_psi}
\end{equation}
while all the other Weyl scalars vanish.  This implies that for
outgoing gravitational waves we can write
\begin{equation}
H = - \int_{-\infty}^t \int_{-\infty}^{t'} \Psi_4 \: dt'' dt' \; . 
\label{eq:HP4}
\end{equation}

\subsection{Radiated energy and momentum}
\label{sec:psi4_radiation}

We can now use equation~\eqref{eq:HP4} to rewrite the expressions for
the radiated energy and momentum derived in the previous Section
directly in term of $\Psi_4$~\footnote{The expressions for the
radiated energy, linear momentum and $z$ component of the angular
momentum in terms of $\Psi_4$ (without the multipole expansion) can
be found in~\cite{Campanelli99}}. However, before doing this it is
convenient to project $\Psi_4$ onto the sphere and describe its
angular dependence in terms of the spin-weighted spherical harmonics
${}_{s}Y^{l,m}$ (see Appendix~\ref{app:harmonics}).  One can easily
show that $\Psi_4$ has spin weight $s=-2$, so that we can expand it as
\begin{equation}
\Psi_4 = \sum_{l=2}^\infty \sum_{m=-l}^l A^{l,m}
\left( {}_{-2}Y^{l,m} (\theta,\phi) \right) \; ,
\label{eq:Psi4_multipole}
\end{equation}
with $A^{l,m}$ the expansion coefficients given by
\begin{equation}
A^{l,m} = \oint \Psi_4 \left( {}_{-2}\bar{Y}^{l,m} (\theta,\phi)
\right) \: d\Omega \; .
\end{equation}
Notice that since we are expanding over the harmonics of spin-weight
$s=-2$, the sum over $l$ starts at $l=2$ (the ${}_{s}Y^{l,m}$ are only
defined for $l \geq |s|$). To simplify notation, from now on we will
drop the limits on the summations and will always assume that the sum
over $l$ starts at 2, while the sum over $m$ goes from $-l$ to $l$.
For summations involving coefficients with indices $l'=l \pm 1$ and
$m'=m \pm 1$ one should only remember that the corresponding
coefficients vanish whenever $l'<2$ and $|m'| > l$.

In order to calculate the total flux of energy leaving the system we
can now use~\eqref{eq:ETT} together with the relation between $H$ and
$\Psi_4$, Eq.~\eqref{eq:HP4}, to find
\begin{equation}
\frac{dE}{dt} = \lim_{r \rightarrow \infty} \frac{r^2}{16\, \pi} \oint
\left| \int_{-\infty}^t \Psi_4 \: d t' \right|^2 d \Omega \; .
\label{eq:energy_Psi4}
\end{equation}
Using now the orthogonality of the ${}_sY^{l,m}$, we can rewrite the
radiated energy as
\begin{equation}
\frac{dE}{dt} = \lim_{r \rightarrow \infty} \frac{r^2}{16\,\pi}
\sum_{l,m} \left| \int_{-\infty}^t A^{l,m} \: dt' \right|^2 \; .
\label{eq:energy_A}
\end{equation}
In a similar way, we can express the linear momentum radiated in terms
of $\Psi_4$ and the $A^{l,m}$ coefficients. Using the expression for
the radiated momentum, Eq.~\eqref{eq:PTT}, and again
Eq.~\eqref{eq:HP4}, we obtain
\begin{equation}
\frac{dP_i}{dt} = \lim_{r \rightarrow \infty} \frac{r^2}{16\, \pi} \oint
l_i \left| \int_{-\infty}^t \Psi_4 \: d t' \right|^2 d \Omega \; .
\label{eq:momentum_Psi4}
\end{equation}
Substituting now the multipole expansion of $\Psi_4$ we find
\begin{eqnarray}
\frac{d P_i}{dt} &=& \lim_{r \rightarrow \infty} \frac{r^2}{16 \pi}
\sum_{l,m} \sum_{l',m'} \oint l_i \Big( {}_{-2}Y^{l,m} \Big)
\Big( {}_{-2}\bar{Y}^{l',m'} \Big) \: d \Omega \nonumber \\
&& \times \int_{-\infty}^t A^{l,m} \: dt'
\int_{-\infty}^t \bar{A}^{l',m'} \: dt' \; .
\end{eqnarray}

In order to calculate the integral over the sphere, notice first that
the components of the radial unit vector $l_i$ can be expressed in
terms of scalar ({{\em i.e.} spin zero) spherical harmonics as
\begin{eqnarray}
l_x &=& \sin \theta \cos \varphi 
= \sqrt{\frac{2 \pi}{3}} \: \Big[ Y^{1,-1} - Y^{1,1} \Big] \; , \\ 
l_y &=& \sin \theta \sin \varphi
= i\,\sqrt{\frac{2 \pi}{3}} \: \Big[ Y^{1,-1} + Y^{1,1} \Big] \; ,
\hspace{5mm} \\
l_z &=& \cos \theta = 2\,\sqrt{\frac{\pi}{3}} \: Y^{1,0} \; .
\end{eqnarray}
We then see that the flux of linear momentum involves integrals over
three spin-weighted spherical harmonics. Such integrals are given in
terms of the Wigner 3-$lm$ symbols with \mbox{$l_3=1$}, and are also
explicitly given in Appendix~\ref{app:harmonics}.

Instead of $P_x$ and $P_y$ it turns out to be easier to work with the
complex quantity $P_+ := P_x + i P_y$. After a straightforward, but
rather long, calculation one finally arrives at the following
expressions for the flux of linear momentum
\begin{eqnarray}
\frac{d P_+}{dt} &=& \lim_{r \to \infty} \frac{r^2}{8\, \pi} \sum_{l,m}
\int_{-\infty}^t  dt' A^{l,m} \nonumber \\
&\times& \int_{-\infty}^t dt' \: \Big( a_{l,m}\, \bar{A}^{l,m+1}
+ b_{l,-m} \,\bar{A}^{l-1,m+1} \nonumber \\
&-&  b_{l+1,m+1}\, \bar{A}^{l+1,m+1} \Big) \; ,
\label{eq:dt_p+} \\
\frac{d P_z}{dt} &=& \lim_{r \to \infty} \frac{r^2}{16 \pi} \sum_{l,m}
\int_{-\infty}^t  dt' A^{l,m} \nonumber \\
&\times& \int_{-\infty}^t dt' \: \Big( c_{l,m}\, \bar{A}^{l,m}
+ d_{l,m}\, \bar{A}^{l-1,m} \nonumber \\
&+&  d_{l+1,m}\, \bar{A}^{l+1,m} \Big) \; ,
\label{eq:dt_pz}
\end{eqnarray}
where the coefficients $(a_{l,m},b_{l,m},c_{l,m},d_{l,m})$ are given by
\begin{eqnarray}
a_{l,m} &=& \frac{\sqrt{(l-m)\,(l+m+1)}}{l\,(l+1)} \; , \\
b_{l,m} &=& \frac{1}{2\,l}\, \sqrt{\frac{(l-2)\,(l+2)\,(l+m)\,(l+m-1)}
{(2l-1)(2l+1)}} \; , \hspace{8mm} \\
c_{l,m} &=& \frac{2\,m}{l\,(l+1)} \; , \\
d_{l,m} &=& \frac{1}{l}\, \sqrt{\frac{(l-2)\,(l+2)\,(l-m)\,(l+m)}
{(2l-1)(2l+1)}} \; .
\end{eqnarray}

Finally, for the flux of angular momentum we go back to
Eq.~\eqref{eq:angularmomentumflux_GW2} to obtain
\begin{eqnarray}
\frac{d J_i}{dt} &=& -\lim_{r \rightarrow \infty}
\frac{r^2}{16 \pi} \: {\rm Re} \Bigg\{ \oint
\left( \int_{-\infty}^t \bar{\Psi}_4 \: d t' \right) \nonumber \\
&& \times \hat{J}_i \left( \int_{-\infty}^t
\int_{-\infty}^{t'} \Psi_4 \: d t'' d t' \right) d \Omega \Bigg\} \; .
\label{eq:angularmomentumflux_GW3}
\end{eqnarray}
Expressing $\Psi_4$ in terms of its multipole expansion and
integrating over the sphere we now find
\begin{eqnarray}
\frac{d J_i}{dt} &=& - \lim_{r \rightarrow \infty}
\frac{r^2}{16\, \pi} \: {\rm Re} \Bigg\{ \sum_{l,m} \sum_{l^\prime m^\prime} 
\int_{-\infty}^t \bar{A}^{l', m'} \: d t'  \nonumber \\ 
&\times& \int_{-\infty}^t \int_{-\infty}^{t'} A^{l,m} \: d t'' d t'
\nonumber \\
&\times& \oint {}_{-2}\bar{Y}^{l', m'}
\hat{J}_i \left( {}_{-2}Y^{l,m} \right) d \Omega \Bigg\} \; ,
\label{eq:angularmomentumflux_GW4}
\end{eqnarray}
where the action of the angular momentum operators $\hat{J}_i$ on the
spin-weighted spherical harmonics is given in
Appendix~\ref{app:harmonics}.  We again obtain integrals that involve
products of two spin-weighted spherical harmonics which satisfy the
usual orthonormalization relations. One can then easily find the
following expressions for the angular momentum carried by the
gravitational waves
\begin{eqnarray}
\frac{d J_x}{dt} &=& - \lim_{r\rightarrow\infty} \frac{i\,r^2}{32 \pi} \:
{\rm Im} \Bigg\{ \sum_{l,m} \int_{-\infty}^t
\int_{-\infty}^{t'} A^{l,m} \: dt'' dt' \hspace{7mm} \nonumber \\
&\times& \int_{-\infty}^t \Big( f_{l,m}\, \bar{A}^{l,m+1}
+ f_{l,-m}\, \bar{A}^{l,m-1} \Big) \Bigg\} dt' ,
\label{eq:dt_jx} \\
\frac{d J_y}{dt} &=& - \lim_{r\rightarrow\infty} \frac{r^2}{32 \pi} \:
{\rm Re} \Bigg\{ \sum_{l,m} \int_{-\infty}^t
\int_{-\infty}^{t'} A^{l,m} \: dt'' dt' \nonumber \\
&\times& \int_{-\infty}^t  \Big( f_{l,m}\, \bar{A}^{l,m+1}
- f_{l,-m}\, \bar{A}^{l,m-1} \Big) \Bigg\} dt' , 
\label{eq:dt_jy} \\
\frac{d J_z}{dt} &=& - \lim_{r\rightarrow\infty} \frac{i r^2}{16 \pi} \:
{\rm Im} \Bigg\{ \sum_{l,m} m \int_{-\infty}^t \int_{-\infty}^{t'} A^{l,m}
\: dt' dt'' \nonumber \\
&\times& \int_{-\infty}^t \bar{A}^{l,m} \: dt' \Bigg\} \; ,
\label{eq:dt_jz}
\end{eqnarray}
with
\begin{eqnarray}
f_{l,m} &:=& \sqrt{(l-m)(l+m+1)} \nonumber \\
&=& \sqrt{l(l+1) - m(m+1)} \; ,
\end{eqnarray}
and where we use the convention that ${\rm Im}(a + i b) = i b$, for
$a$ and $b$ real. These last expressions have also been recently
derived following a different route by Lousto and Zlochower
in~\cite{Lousto:2007mh}. The different factor of $1/4$ between our
expression and the expressions of~\cite{Lousto:2007mh} is due to a
different normalization of the null tetrad used to define $\Psi_4$.


\section{black hole perturbation theory}
\label{sec:GW_perturbations}

In order to relate the expressions for the radiated energy and
momentum in terms of the Weyl scalar $\Psi_4$ to the standard ones in
terms of gauge invariant perturbations, we will present here a brief
discussion of some of the ideas behind the theory of perturbations of
a Schwarzschild black hole (a more detailed discussion can be seen in
{\em e.g.}~\cite{Nagar:2005ea,Martel:2005ir,Sarbach:2001qq}).

\subsection{Multipole expansion}
\label{GW_multipoles}

Consider a metric of the form~\eqref{eq:pertubation}, with the
background metric $\gz_{\mu \nu}$ given by the Schwarzschild metric in
standard coordinates:
\begin{equation}
\gz_{\mu \nu} \: dx^\mu dx^\nu = f(r) \: dt^2 + \frac{1}{f(r)} \: dr^2
+ r^2\, d\Omega^2 \; ,
\end{equation}
with $f(r) = (1 - 2M/r)$. Because of the spherical symmetry of the
background, it is convenient to think of the full spacetime as the
product of a Lorentzian two-dimensional manifold $M^2$ associated to
the coordinates $(t,r)$, and the two-sphere of unit radius $S^2$
associated to $(\theta,\varphi)$:
\begin{equation}
ds^2 = g_{AB} \: dx^A dx^B + r^2 \Omega_{ab} \: dx^a dx^b \; ,
\label{eq:split}
\end{equation}
where $\Omega_{ab}$ is the metric on $S^2$: $\Omega_{ab} = {\rm
diag}(1,\sin^2 \theta)$. Here and in what follows we will use upper
case indices $(A,B,\dots)$ to represent the coordinates in $M^2$, and
lower case indices $(a,b,\dots)$ for coordinates in $S^2$. We will
also distinguish covariant derivatives in the full spacetime from
covariant derivatives in the submanifolds: $\nabla_\mu$ will represent
covariant derivatives in spacetime, while $D_A$ and $D_a$ will denote
covariant derivatives in $M^2$ and $S^2$ respectively.

We will now consider an expansion of the metric perturbation $h_{\mu
\nu}$ in multipoles using spherical harmonics
$Y^{l,m}(\theta,\varphi)$.  Such a decomposition naturally separates
the perturbation into even (or axial) modes and odd (or polar) modes:
even modes are those that transform as $(-1)^l$ under a parity
transformation $(\theta,\phi)\rightarrow\,(\pi-\theta,\pi+\phi)$,
while odd modes transform instead as $(-1)^{l+1}$.

In order to decompose $h_{\mu \nu}$ one further needs to introduce the
scalar, vector and tensor harmonics (these tensorial properties refer
only to transformations in the unit sphere). Scalar harmonics are the
usual functions $Y^{l,m}$.  Vector harmonics, on the other hand, come
in two different types.  The even vector harmonics are simply defined
as the gradient of the scalar harmonics on the sphere
\begin{equation}
Y^{l,m}_a := D_a Y^{l,m} \; ,
\end{equation}
while the odd vector harmonics are
\begin{equation}
X^{l,m}_a := - {\epsilon_a}^b D_b Y^{l,m}
= - \epsilon_{ac} \: \Omega^{cb} D_b Y^{l,m} \; ,
\end{equation}
where $\epsilon_{ab}$ is the Levi--Civita tensor on the two-sphere
($\epsilon_{\theta \varphi} = - \epsilon_{\varphi \theta} =
\Omega^{1/2} = \sin \theta$).

Similarly, one can define tensor harmonics of even and odd type.  Even
tensor harmonics can be constructed in two ways, either by multiplying
the scalar harmonics with the angular metric $\Omega_{ab}$, or by
taking a second covariant derivative of the $Y^{l,m}$. However, it
turns out that these functions do not form a linearly independent set.
Instead of the $D_a D_b Y^{l,m}$ it is better to use the so-called
``Zerilli--Mathews'' tensor harmonics (see {\em e.g.}~\cite{Nagar:2005ea})
defined as
\begin{equation}
Z^{l,m}_{ab} := D_a D_b Y^{l,m} + \frac{1}{2} \: l \left( l+1 \right)
\Omega_{ab} Y^{l,m} \;.
\label{eq:Zab}
\end{equation}
It is easy to show that this tensor is traceless, which implies that 
the resultant functions are  linearly independent. 
Odd parity tensor harmonics, on the other hand, can be
constructed in only one way, namely
\begin{equation}
X^{l,m}_{ab} = \frac{1}{2} \left( D_a X_b^{l,m} + D_b X_a^{l,m} \right) \; .
\label{eq:Xab}
\end{equation}

Having defined the vector and tensor harmonics, the perturbed metric
is expanded in multipoles, and separated into its even sector given by
\begin{eqnarray}
\left( h_{AB}^{l,m} \right)_{\rm even} &=& H_{AB}^{l,m} \: Y^{l,m} \; , \\
\left( h_{Ab}^{l,m} \right)_{\rm even} &=& H_A^{l,m} \: Y_b^{l,m} \; , \\
\hspace{-2mm}\left( h_{ab}^{l,m} \right)_{\rm even} &=& r^2 \left( K^{l,m}
\: \Omega_{ab} Y^{l,m} + G^{l,m} Z^{l,m}_{ab} \right) \;, \hspace{5mm}
\end{eqnarray}
notice that for the  case  $l=1$ the tensor  $Z^{l,m}_{ab}$ vanishes,
and  the odd sector is given by
\begin{eqnarray}
\left( h_{AB}^{l,m} \right)_{\rm odd} &=& 0 \; , \\
\left( h_{Ab}^{l,m} \right)_{\rm odd} &=& h_A^{l,m} \: X_b^{l,m} \; , \\
\left( h_{ab}^{l,m} \right)_{\rm odd} &=& h^{l,m} \: X^{l,m}_{ab} \; ,
\end{eqnarray}
where the coefficients $(H_{AB}^{l,m}, H_A^{l,m}, K^{l,m}, G^{l,m},
h_A^{l,m}, h^{l,m})$ are in general functions of $r$ and $t$.

Notice that, since $Y^{00}$ is a constant, both vector and tensor
harmonics vanish for $l=0$.  On the other hand, for $l=1$ the vector
harmonics do not vanish, but the tensor harmonics can still be easily
shown to vanish from the explicit expressions for $Y^{1,m}$.  This
means that vector harmonics are only non-zero for $l \geq 1$, and
tensor harmonics for $l \geq 2$. The scalar mode with $l=0$ can be
interpreted as a variation in the mass of the Schwarzschild spacetime,
while the even mode with $l=1$ is just gauge and can be removed under
a suitable
transformation~\cite{Gleiser:1999hw,Sarbach:2001qq,Martel:2005ir}. On
the other hand, the odd mode with $l=1$ can be interpreted as an
infinitesimal angular momentum contribution, {\em i.e.} a ``Kerr''
mode (a detailed discussion about this point can be found
in~\cite{Sarbach:2001qq}). We will therefore only be interested in
modes with $l \geq 2$, just as it happened with the expansion of
$\Psi_4$.


\subsection{Gauge invariant perturbations}
\label{sec:GW_gaugeinvariant}

As we have seen, even parity perturbations are characterized by the
coefficients $(H^{l,m}_{AB}, H^{l,m}_A, K^{l,m}, G^{l,m})$.  These
coefficients are clearly coordinate dependent, and in particular
change under infinitesimal coordinate transformations of the form
\mbox{$x^\mu \rightarrow x^\mu + \xi^\mu$}, with $|\xi^\mu| \ll 1$.
However, it turns out that one can construct gauge invariant
combinations of coefficients. Two such invariant combinations are (for
details see {\em e.g.}~\cite{Gerlach79,Martel:2005ir,Nagar:2005ea})
\begin{eqnarray}
\tilde{K}^{l,m} &:=& K^{l,m} + \frac{1}{2} \: l (l+1) \: G^{l,m}
- \frac{2}{r} \: r^A \varepsilon_A^{l,m} \; , \hspace{5mm} \\
\tilde{H}^{l,m}_{AB} &:=& H^{l,m}_{AB} - D_A \varepsilon^{l,m}_B
- D_B \varepsilon^{l,m}_A \; ,
\end{eqnarray}
with $\varepsilon^{l,m}_A := H^{l,m}_A - r^2 D_A G^{l,m}/2$.  In terms of
the gauge invariant perturbations $\tilde{K}^{l,m}$ and
$\tilde{H}^{l,m}_{AB}$ one defines the so-called ``Zerilli--Moncrief
master function'' as
\begin{eqnarray}
\Psi^{l,m}_{\rm even} &:=& \frac{2\,r}{L} \left[
\tilde{K}^{l,m} \right. \nonumber \\
&+& \left. \frac{2\, r^A}{\Lambda} \left( r^B\, \tilde{H}^{l,m}_{AB}
- r\, D_A \tilde{K}^{l,m} \right) \right] , \hspace{4mm}
\label{eq:Psi_even}
\end{eqnarray}
where $L:= l(l+1)$, $\Lambda := (l-1)(l+2) + 6M/r$, and $r_A := D_A
r$. This quantity can be shown to obey a simple wavelike equation
known as the ``Zerilli equation''~\cite{Zerilli70}, though we will not
go into such details here.

One can also construct a gauge invariant quantity for the case of odd
perturbations in the following way (again, see {\em
e.g.}~\cite{Gerlach79,Martel:2005ir,Nagar:2005ea})
\begin{equation}
\tilde{h}_A^{l,m} := h_A^{l,m} - \frac{1}{2}\,r^2 \, D_A
\left( \frac{h^{l,m}}{r^2} \right) \; .
\end{equation}
In terms of $\tilde{h}_A^{l,m}$ one now defines the
``Cunningham--Price--Moncrief master function'' as
\begin{eqnarray}
\Psi^{l,m}_{\rm odd} &:=&
\frac{2\,r \, \epsilon^{AB}}{(l-1)\,(l+2)} \left[ D_A \tilde{h}^{l,m}_B
- \frac{2\, r_A}{r} \, \tilde{h}^{l,m}_B \right]
\nonumber \\
&=& \frac{2\,r \, \epsilon^{AB}}{(l-1)\,(l+2)} \left[ D_A h^{l,m}_B
- \frac{2\, r_A}{r} \, h^{l,m}_B \right] \: . \hspace{8mm} 
\label{eq:Psi_odd}
\end{eqnarray}
The second equality shows that $\Psi_{\rm odd}$ in fact depends only
on $h_A$ and not on $h$ (the contributions from $h$ cancel when
contracted with the $\epsilon^{AB}$), but it is nevertheless gauge
invariant. Again, using the perturbed vacuum Einstein field equations
one can show that $\Psi_{\rm odd}$ obeys a simple wavelike equation
known as the ``Regge--Wheeler equation''~\cite{Regge57}.


\subsection{Gravitational radiation in the TT gauge}
\label{sec:grav_TT}

The advantage of using the gauge invariant quantities introduced above
is that they can be naturally related to gravitational waves in the TT
gauge.

Consider first the even parity perturbations.  If asymptotically we
approach the TT gauge then we will find that $h_{AB}$ and $h_{Ab}$
decay much faster than $h_{ab}$. According to the multiple expansion
we can then ignore the coefficients $H^{l,m}_{AB}$ and $H^{l,m}_A$, and
concentrate only on $K^{l,m}$ and $G^{l,m}$. Considering a local
orthonormal basis aligned with the angular directions one then finds
that
\begin{eqnarray}
(h^+)^{l,m}_{\rm even} &=& \frac{G^{l,m}}{2} \left( Z^{l,m}_{\theta \theta}
- \frac{Z^{l,m}_{\varphi \varphi}}{\sin^2 \theta} \right) \; , \\
(h^\times)^{l,m}_{\rm even} &=& G^{l,m} \left(
\frac{Z^{l,m}_{\theta \varphi}}{\sin \theta} \right) \; .
\end{eqnarray}
On the other hand, from the traceless condition one finds that
$K=0$. In that case the Zerilli--Moncrief function~\eqref{eq:Psi_even}
simplifies to \mbox{$\Psi^{l,m}_{\rm even} = r\,G^{l,m}$}, which implies
that the contribution from even perturbations to the TT metric
functions can be written in terms of $\Psi_{\rm even}$ as
\begin{eqnarray}
(h^+)^{l,m}_{\rm even} &=& \frac{\Psi^{l,m}_{\rm even}}{2\,r}
\left( Z^{l,m}_{\theta \theta}
- \frac{Z^{l,m}_{\varphi \varphi}}{\sin^2 \theta} \right) \nonumber \\
&=& \frac{\Psi^{l,m}_{\rm even}}{r} \left[ \frac{\partial^2}{\partial \theta^2} 
+ \frac{1}{2} \: l (l+1) \right] Y^{l,m} \; , \\
(h^\times)^{l,m}_{\rm even} &=& \frac{\Psi^{l,m}_{\rm even}}{r} \:
\left( \frac{Z^{l,m}_{\theta \varphi}}{\sin \theta} \right) \nonumber \\
&=& \frac{\Psi^{l,m}_{\rm even}}{r} \left( \frac{i\,m}{\sin \theta} \right)
\left[ \frac{\partial}{\partial \theta} - \cot \theta \right] Y^{l,m} ,
\hspace{5mm}
\end{eqnarray}
where we used the fact that $\partial_\varphi Y^{l,m} = i\,m\,Y^{l,m}$.

Consider now the odd perturbations.  We find,
\begin{eqnarray}
(h^+)^{l,m}_{\rm odd} &=& \frac{h^{l,m}}{2\,r^2} \left( X^{l,m}_{\theta \theta}
- \frac{X^{l,m}_{\varphi \varphi}}{\sin^2 \theta} \right) \; , \\
(h^\times)^{l,m}_{\rm odd} &=& \frac{h^{l,m}}{r^2} \left(
\frac{X^{l,m}_{\theta \varphi}}{\sin \theta} \right) \; .
\end{eqnarray}
We now need to relate $h^{l,m}$ to $\Psi^{l,m}_{\rm odd}$.  In this
case, however, we cannot just ignore $h^{l,m}_A$ in favor of
$h^{l,m}$, since from the definition of $\Psi^{l,m}_{\rm odd}$ we see
that it in fact depends only on $h^{l,m}_A$ and not on
$h^{l,m}$. Nevertheless, in the TT gauge these quantities are related
to each other.  In order to see this, consider the transverse
condition on $h_{\mu a}$, $\nabla^\mu h_{\mu a} = 0$.  Using the
multipole expansion, and calculating explicitly the divergence of
$X^{l,m}_{ab}$, one finds that this condition implies
\begin{equation}
D^A \left( r^2\,h^{l,m}_A \right)
= \frac{1}{2} \: (l-1)(l+2) \: h^{l,m} \; .
\end{equation}
Remembering now that in the TT gauge one has the extra freedom of
taking $h_{\mu \nu}$ to be purely spatial, plus the fact that
asymptotically the metric $g_{AB}$ is just the Minkowski metric, the
previous expression reduces to
\begin{equation}
\partial_r \left( r^2 h^{l,m}_r \right)
= \frac{1}{2} \: (l-1)(l+2) \: h^{l,m} \; .
\label{eq:transverse_odd}
\end{equation}
In the same limit one can also rewrite expression~\eqref{eq:Psi_odd}
for $\Psi^{l,m}_{\rm odd}$ as
\begin{equation}
\Psi^{l,m}_{\rm odd} = \frac{2\,r}{(l-1)\,(l+2)} \: \partial_t h^{l,m}_r \; .
\end{equation}
Collecting results we find that
\begin{equation}
\partial_r \left( r\, \Psi^{l,m}_{\rm odd} \right)
= \partial_t h^{l,m} \; .
\end{equation}
Since for an outgoing wave we have $\partial_t h^{l,m}_r \sim -
\partial_r h^{l,m}_r$, we can integrate the above expression to find
\begin{equation}
h^{l,m} \sim -  r\,\Psi^{l,m}_{\rm odd} \; .
\end{equation}
We can then rewrite the odd metric perturbations as
\begin{eqnarray}
(h^+)^{l,m}_{\rm odd} &=& - \frac{\Psi^{l,m}_{\rm odd}}{2 r}
\left( X^{l,m}_{\theta \theta}
- \frac{X^{l,m}_{\varphi \varphi}}{\sin^2 \theta} \right) \nonumber \\
&=& - \frac{\Psi^{l,m}_{\rm odd}}{r} \left( \frac{im}{\sin \theta} \right)
\left[ \frac{\partial}{\partial \theta} - \cot \theta \right] Y^{l,m} ,
\hspace{5mm} \\
(h^\times)^{l,m}_{\rm odd} &=& - \frac{\Psi^{l,m}_{\rm odd}}{r} \left(
\frac{X^{l,m}_{\theta \varphi}}{\sin \theta} \right) \nonumber \\
&=& \frac{\Psi^{l,m}_{\rm odd}}{r} \left[ \frac{\partial^2}{\partial \theta^2} 
+ \frac{1}{2} \: l (l+1) \right] Y^{l,m} .
\end{eqnarray}
The full TT coefficients $h^+$ and $h^\times$ then take the
form
\begin{eqnarray}
h^+ &=& \frac{1}{2\,r} \: \sum_{l,m} \left[ \Psi^{l,m}_{\rm even}
\left( Z^{l,m}_{\theta \theta}
- \frac{Z^{l,m}_{\varphi \varphi}}{\sin^2 \theta} \right) \right. \nonumber \\
&& \left. - \Psi^{l,m}_{\rm odd} \left( X^{l,m}_{\theta \theta}
- \frac{X^{l,m}_{\varphi \varphi}}{\sin^2 \theta} \right)
\right] \: , \hspace{8mm} \\
h^\times &=& \frac{1}{r} \: \sum_{l,m} \left[ \Psi^{l,m}_{\rm even}
\left( \frac{Z^{l,m}_{\theta \varphi}}{\sin \theta} \right) \right.
\nonumber \\
&& \left. - \Psi^{l,m}_{\rm odd}
\left( \frac{X^{l,m}_{\theta \varphi}}{\sin \theta}\right) \right] \: .
\end{eqnarray}

Notice that while the functions $(\Psi^{l,m}_{\rm even},\Psi^{l,m}_{\rm
even})$ are in general complex, the TT coefficients $h^+$ and
$h^\times$ must be real. Using the properties of the spherical
harmonics under complex conjugation one can easily show that this
implies
\begin{equation}
\bar{\Psi}^{l,m}_{\rm even} = (-1)^m \: \Psi^{l,-m}_{\rm even} \; , \quad
\bar{\Psi}^{l,m}_{\rm odd} = (-1)^m \: \Psi^{l,-m}_{\rm odd} \; .
\label{eq:Psi_complex}
\end{equation}

One can also rewrite the expressions above in terms of spin-weighted
spherical harmonics.  In order to do this it is in fact more
convenient to consider the complex combination $H$, for which we find
\begin{equation}
H = \frac{1}{2\,r} \sum_{l,m} \sqrt{\frac{(l+2)!}{(l-2)!}} \:
\Big( \Psi^{ }_{\rm even} + i\, \Psi^{ }_{\rm odd} \Big) \:
{}_{-2}Y^{l,m} .
\label{eq:H_Psis}
\end{equation}

At this point it is important to mention one very common convention
used in numerical relativity. We start by considering a different odd
master function originally introduced by Moncrief~\cite{Moncrief74}:
\begin{eqnarray}
Q^{l,m}_{\rm M} &:=& \frac{2 r^A \tilde{h}^{l,m}_A}{r} \nonumber \\
&=& \frac{2\, r^A}{r} \: \left( h_A^{l,m} - \frac{1}{2} \: D_A h^{l,m} +
\frac{r_A}{r} \: h^{l,m} \right) . \hspace{5mm}
\end{eqnarray}
Notice that with this definition $Q^{l,m}_{\rm M}$ is clearly scalar
and gauge invariant.  The Moncrief function just defined has
traditionally been the most common choice to study odd perturbations
of Schwarzschild, and because of this many numerical implementations
use this function instead of $\Psi^{l,m}_{\rm odd}$.  It is possible to
show that asymptotically, and in the TT gauge, ${Q}^{l,m }_{\rm M}$
reduces to
\begin{equation}
Q^{l,m}_{\rm M} \sim - \partial_t \Psi^{l,m}_{\rm odd} \; .
\end{equation}
One finally introduces the following rescaling
\begin{eqnarray}
Q^{l,m}_{\rm even} &:=& \sqrt{\frac{(l+2)!}{2(l-2)!}}\,
\Psi^{l,m}_{\rm even} \; , \\
Q^{l,m}_{\rm odd} &:=& \sqrt{\frac{(l+2)!}{2(l-2)!}}\,
Q^{l,m}_{\rm M} \; .
\end{eqnarray}
In terms of $Q^{l,m}_{\rm even}$ and $Q^{l,m}_{\rm odd}$, one now finds
for the complex quantity $H$:
\begin{equation}
H = \frac{1}{\sqrt{2}\,r} \: \sum_{l,m}
\left[ Q^{l,m}_{\rm even} - i \int_{-\infty}^t Q^{l,m}_{\rm odd} \: dt'
\right] \: {}_{-2}Y^{l,m} \; .
\label{eq:H_Qs}
\end{equation}


\subsection{Radiated energy and momentum}

Just as we did for the case of $\Psi_4$, we can now express the
radiated energy, linear momentum and angular momentum in terms of the
master functions $(\Psi^{l,m}_{\rm even}, \Psi^{l,m}_{\rm odd})$ and/or
$(Q^{l,m}_{\rm even},Q^{l,m}_{\rm odd})$.  There are two ways in which
one can do these calculations.  One approach is to substitute directly
equations~\eqref{eq:H_Psis} and~\eqref{eq:H_Qs} into the expressions
for the radiated energy and momentum. It is much easier, however, to
start from the expressions for the radiated energy and momentum in
terms of the multipole expansion of $\Psi_4$, and relate the
coefficients $A^{l,m}$ to $(\Psi^{l,m}_{\rm even}, \Psi^{l,m}_{\rm odd})$
and $(Q^{l,m}_{\rm even},Q^{l,m}_{\rm odd})$.

Comparing the multipole expansion for $\Psi_4$ with the expansions for
the metric perturbations, and using the fact asymptotically
\mbox{$\Psi_4 = - \ddot{H}$}, we find
\begin{eqnarray}
A^{l,m} &=& - \frac{1}{2\,r}\, \sqrt{\frac{(l+2)!}{(l-2)!}}\,
\left( \ddot{\Psi}^{l,m}_{\rm even} + i\,\ddot{\Psi}^{l,m}_{\rm odd}
\right) \nonumber \\
&=& - \frac{1}{\sqrt{2} \: r} \left( \ddot{Q}^{l,m}_{\rm even}
- i\, \dot{Q}^{l,m}_{\rm odd} \right) \:.
\label{eq:AfromQ}  
\end{eqnarray}
Using this one can translate expressions in terms of the $A^{l,m}$
directly into expressions in terms of $(\Psi^{l,m}_{\rm even},
\Psi^{l,m}_{\rm odd})$ and/or $(Q^{l,m}_{\rm even},Q^{l,m}_{\rm odd})$.

Let us consider first the radiated energy. Using
equations~\eqref{eq:energy_A} and~\eqref{eq:AfromQ}, one immediately
obtains
\begin{eqnarray}
\frac{dE}{dt} &=& \frac{1}{64 \,\pi}
\sum_{l,m} \frac{(l+2)!}{(l-2)!}
\left( \left| \dot{\Psi}^{l,m}_{\rm even} \right|^2
+ \left| \dot{\Psi}^{l,m}_{\rm odd} \right|^2 \right) \nonumber \\
&=& \frac{1}{32\,\pi} \sum_{l,m}
\left( \left| \dot{Q}^{l,m}_{\rm even} \right|^2
+ \left| \rule{0mm}{4mm} Q^{l,m}_{\rm odd} \right|^2 \right) \; ,
\label{eq:energyflux_multipole2}
\end{eqnarray}
where, in order to derive these expressions one must use the fact that,
as a consequence of~\eqref{eq:Psi_complex},
\begin{equation}
\sum_m \left( \dot{\Psi}^{l,m}_{\rm even}\, \dot{\bar{\Psi}}^{l,m}_{\rm odd}
- \dot{\bar{\Psi}}^{l,m}_{\rm even}\, \dot{\Psi}^{l,m}_{\rm odd} \right)
= 0 \; .
\end{equation}

For the linear momentum we start from equations
\eqref{eq:dt_p+}-\eqref{eq:dt_pz}, and again use~\eqref{eq:AfromQ}.
The calculation is now considerably longer, and in order to simplify
the expressions one must make use several times
of~\eqref{eq:Psi_complex} (or its equivalent in terms of the
$Q$'s). The final result is
\begin{eqnarray}
\frac{d P_+}{dt} &=& \frac{1}{8 \pi} \sum_{l,m} \left[
i \, a_{l,m} \, \dot{Q}^{l,m}_{\rm even}\, \bar{Q}^{l,m+1}_{\rm odd} \right.
\nonumber \\
&& \left. \hspace{-12mm}
-b_{l+1,m+1} \left( \dot{Q}^{l,m}_{\rm even} \dot{\bar{Q}}^{l+1,m+1}_{\rm even}
+ Q^{l,m}_{\rm odd} \,\bar{Q}^{l+1,m+1}_{\rm odd} \right) \right] , \hspace{8mm}
\label{eq:p+_q} \\
\frac{d P_z}{dt} &=& \frac{1}{16 \pi} \sum_{l,m} \left[
i\, c_{l,m} \, \dot{Q}^{l,m}_{\rm even}\, \bar{Q}^{l,m}_{\rm odd} \right.
\nonumber \\
&& \left. \hspace{-12mm}
+ d_{l+1,m} \left( \dot{Q}^{l,m}_{\rm even}\, \dot{\bar{Q}}^{l+1,m}_{\rm even}
+ Q^{l,m}_{\rm odd} \,\bar{Q}^{l+1,m}_{\rm odd} \right) \right] ,
\label{eq:pz_q}
\end{eqnarray}
with the coefficients $(a_{l,m},b_{l,m},c_{l,m},d_{l,m})$ the same as
before.  The last expressions are written only in terms of
$(Q^{l,m}_{\rm even},Q^{l,m}_{\rm odd})$, but it is trivial to rewrite
them in terms of $(\Psi^{l,m}_{\rm even},\Psi^{l,m}_{\rm odd})$.
These expressions can be easily shown to be equivalent to those
recently derived by Pollney {\em et al.} in~\cite{Pollney:2007ss},
and by Sopuerta {\em et al.} in~\cite{Sopuerta:2006wj},
(but one must be careful when comparing with the first reference,
as their sums over $m$ go only from 0 to $l$).

In a similar way we can obtain expressions for the radiated angular
momentum staring from equations~\eqref{eq:dt_jx}-\eqref{eq:dt_jz}.
One now finds, after some algebra, 
\begin{eqnarray}
\frac{d J_x}{dt} &=& \frac{i}{32 \pi} \: {\rm Im} \sum_{l,m} 
f_{l,m} \left( \bar{Q}^{l,m}_{\rm even}\, \dot{Q}^{l,m+1}_{\rm even}
\right. \nonumber \\
&& \left. \hspace{5mm}
+ \bar{P}^{l,m}_{\rm odd}\, Q^{l,m+1}_{\rm odd} \right) \: ,
\label{eq:jx_q} \\
\frac{d J_y}{dt} &=& - \frac{1}{32 \pi} \: {\rm Re} \sum_{l,m}
f_{l,m} \left( \bar{Q}^{l,m}_{\rm even} \,\dot{Q}^{l,m+1}_{\rm even}
\right. \nonumber \\
&& \left. \hspace{5mm}
+ \bar{P}^{l,m}_{\rm odd} Q^{l,m+1}_{\rm odd} \right) \: ,
\label{eq:jy_q} \\
\frac{d J_z}{dt} &=& \frac{i}{32 \pi} \sum_{l,m}
m \left( \dot{Q}^{l,m}_{\rm even}\, \bar{Q}^{l,m}_{\rm even} 
+ Q^{l,m}_{\rm odd}\, \bar{P}^{l,m}_{\rm odd} \right) \: , \hspace{8mm}
\label{eq:jz_q}
\end{eqnarray}
where we have defined
\begin{equation}
P^{l,m}_{\rm odd} := \int^t_{-\infty}{Q}^{l,m}_{\rm odd} \: dt' \; ,
\end{equation}
and where again the coefficients $f_{l,m}$ are the same as before.
Notice that the expressions for $dJ_x/dt$ and $dJ_y/dt$ are manifestly
real. For $dJ_z/dt$ the term inside the sum can be easily shown to be
purely imaginary, so that the final result is also real.

Equivalent expressions to the set of equations~\eqref{eq:energy_A},
\eqref{eq:dt_p+}-\eqref{eq:dt_pz}
and~\eqref{eq:dt_jx}-\eqref{eq:dt_jz} for the energy and momenta
carried away by gravitational waves in terms of $\Psi_4$, or the the
set~\eqref{eq:energyflux_multipole2}, \eqref{eq:p+_q}-\eqref{eq:pz_q}
and~\eqref{eq:jx_q}-\eqref{eq:jz_q} in terms of gauge invariant
perturbations, were derived by Thorne in~\cite{Thorne80b}. One can
directly relate them to the results presented here by noticing that
\begin{eqnarray}
A^{l,m} &=& \frac{1}{\sqrt{2}\,r} \left[ ^{(l+2)}I^{l,m}
- i \left( {} ^{(l+2)}S^{l,m} \right) \right] \; , \nonumber \\
\bar{A}^{l,m} &=& \frac{(-1)^m}{\sqrt{2}\,r} \left[ ^{(l+2)}I^{l,-m}
+ i \left( {}^{(l+2)}S^{l,-m} \right) \right] \; , \hspace{8mm}
\label{eq:Thorne}
\end{eqnarray}
where in Thorne's notation $I^{l,m}(t-r)$ is the {\em mass multipole
momenta}\/ of the radiation field, $S^{l,m}(t-r)$ is the {\em current
multipole momenta}\/, and where $^{(l)}I^{l,m}$ and $^{(l)}S^{l,m}$
denote the $l$th time derivative of these quantities.

As final comment, notice that in order to simplify the notation in
this Section we have not explicitly introduced the limit of infinite
radius. Nevertheless, this limit should be understood since all the
results are valid only in the weak field approximation.


\section{Conclusions}
\label{sec:conclusions}

In this paper we have considered explicit expressions for the energy,
linear momentum and angular momentum radiated by an isolated system in
the form of gravitational waves.  Starting from a small perturbation
$h_{\mu\nu}$ a background metric $\gz_{\mu\nu}$, and working in the
transverse-traceless gauge, we have reviewed the standard expressions
for the radiated energy and momentum based on the Isaacson
stress-energy tensor.  Introducing the Weyl scalar $\Psi_4$ and its
multipole expansion in terms of spin-weighted spherical harmonics, we
have computed explicit expressions for the radiated energy and
momentum in terms of the expansion coefficients $A^{l,m}$.  Finally,
we have also presented multipole expansions in terms of the gauge
invariant perturbations of a Schwarzschild spacetime. In particular,
the expressions in terms of the expansion of the Weyl scalar $\Psi_4$
have the advantage of avoiding the need of separating out a
Schwarzschild ``background'' in standard coordinates from a
numerically generated spacetime that can be in an arbitrary gauge.
Nevertheless, one still needs to find a suitable tetrad to calculate
the scalar $\Psi_4$, and as yet there is no standard procedure to find
it (though some recent progress has been made on this issue, see {\em
e.g.}~\cite{Nerozzi:2007ai}).  However, as long as one chooses a
tetrad that approaches the standard outgoing null tetrad in flat space
for large $r$ the asymptotic expressions will be correct.

Although most of the expressions derived here are known, as far as we
know they have never appeared together in the literature. We have
taken great care of explaining the origin of all relevant expressions,
using clear and consistent conventions and definitions. We believe
that having all these expressions together will be extremely useful
when calculating radiated energy and momenta from numerical
simulations of astrophysical systems. As a final comment, we have
available a Mathematica script to find the energy and momenta carried
by gravitational waves, and would be happy to provide it to interested
readers upon request.


\begin{acknowledgments}

We would like to thank Olivier Sarbach for useful comments and discussions.
This work was supported in part by Direcci\'on General de Estudios de
Posgrado (DGEP-UNAM), by CONACyT through grants 47201-F, 47209-F, and by
DGAPA-UNAM through grant IN113907.

\end{acknowledgments}


\appendix


\section{Spin-Weighted Spherical Harmonics}
\label{app:harmonics}

In this appendix we will discuss some important properties of the
spin-weighted spherical harmonics.  Spin-weighted spherical harmonics
where first introduced by Newman and Penrose~\cite{Newman66} for the
study of gravitational radiation, but they can also be used to study
solutions of the Maxwell equations, the Dirac equation, or in fact
dynamical equations for fields of arbitrary spin.

Consider a complex function $f$ on the sphere that might correspond to
some combination of components of a tensorial (or spinorial) object in
the orthonormal basis $(\hat{e}_r,\hat{e}_\theta,\hat{e}_\varphi)$
induced by the spherical coordinates $(r,\theta,\varphi)$. We will say
that $f$ has spin weight $s$ if, under a rotation of the angular basis
$(\hat{e}_\theta,\hat{e}_\varphi)$ by an angle $\psi$, it transforms
as $f \rightarrow e^{- i s \psi} f$.  A trivial example is a scalar
function whose spin weight is clearly zero.  A more interesting
example corresponds to a three-dimensional vector $\vec{v}$ with
components $(v^{\hat{r}},v^{\hat{\theta}},v^{\hat{\phi}})$.  Notice
that these components are different from those in the coordinate basis
(which is not orthonormal), and are related to them through
$(v^{\hat{r}},v^{\hat{\theta}},v^{\hat{\phi}}) = (v^r,r v^\theta,r
\sin \theta \: v^\phi)$ .  Define now two unit complex vectors as
\begin{equation}
\hat{e}_\pm := \left( \hat{e}_\theta
\mp i \hat{e}_\varphi \right) / \sqrt{2} \; .
\end{equation}
The vector $\vec{v}$ can then be written as
\begin{equation}
\vec{v} = v^0 \hat{e}_r + v^+ \hat{e}_+ + v^{-} \hat{e}_{-} \; ,
\end{equation}
where $v^0 := v^{\hat{r}}$, $v^{\pm} := ( v^{\hat{\theta}} \pm i
v^{\hat{\phi}} ) / \sqrt{2}$. By considering a rotation of the vectors
$(\hat{e}_\theta,\hat{e}_\varphi)$ by an angle $\psi$ it is now easy
to see that $v^0$ has spin weight zero, while the spin weight of
$v^{\pm}$ is $\pm 1$.

The spin-weighted spherical harmonics, denoted by ${}_s\bar{Y}^{l,m}
(\theta,\varphi)$, form a basis for the space of functions with
definite spin weight $s$.  They can be introduced in a number of
different ways. One can start by defining the operators
\begin{eqnarray}
\eth f &:=& - \sin^{s} \theta \left( \partial_\theta + \frac{i}{\sin
\theta} \: \partial_\varphi \right) \left( f \sin^{-s} \theta \right)
\nonumber \\
&=& - \left( \partial_\theta + \frac{i}{\sin
\theta} \: \partial_\varphi - s \cot \theta \right) f \; , \\
\bar{\eth} f &:=& - \sin^{-s} \theta \left( \partial_\theta -
\frac{i}{\sin \theta} \: \partial_\varphi \right) \left( f \sin^{s}
\theta \right) \nonumber \\
&=& - \left( \partial_\theta - \frac{i}{\sin
\theta} \: \partial_\varphi + s \cot \theta \right) f \; ,
\end{eqnarray}
where $s$ is the spin weight of $f$. The spin-weighted spherical
harmonics are then defined for $|m| \leq l$ and $l \geq |s|$ in terms
of the standard spherical harmonics as
\begin{eqnarray}
{}_sY^{l,m} := \left[ \frac{(l-s)!}{(l+s)!} \right]^{1/2}
\eth^s \left( Y^{l,m} \right) , \hspace{6mm} && s \geq 0 , \\
{}_sY^{l,m} := (-1)^s \left[ \frac{(l+s)!}{(l-s)!}
\right]^{1/2} \bar{\eth}^{-s} \left( Y^{l,m} \right) ,
&& s \leq 0 . \hspace{8mm}
\end{eqnarray}
In particular we have ${}_0Y^{l,m} = Y^{l,m}$.  The above definition
implies that
\begin{eqnarray}
\eth \left( {}_sY^{l,m} \right) &=& + \left[(l-s)(l+s+1)\right]^{1/2}
{}_{s+1}Y^{l,m} \; , \hspace{5mm} \\
\bar{\eth} \left( {}_sY^{l,m} \right) &=& - \left[(l+s)(l-s+1)\right]^{1/2}
{}_{s-1}Y^{l,m} \; .
\end{eqnarray}
Because of this, $\eth$ and $\bar{\eth}$ are known as the {\em spin
raising} and {\em spin lowering} operators.  One also finds that
\begin{eqnarray}
\bar{\eth} \eth \left( {}_sY^{l,m} \right) &=& - \left[ l(l+1)
- s(s+1) \right] \: {}_sY^{l,m} \; , \hspace{5mm} \\
\eth \bar{\eth} \left( {}_sY^{l,m} \right) &=& - \left[ l(l+1)
- s(s-1) \right] \: {}_sY^{l,m} \; ,
\end{eqnarray}
so the ${}_sY^{l,m}$ are eigenfunctions of the operators $\bar{\eth}
\eth$ and $\eth \bar{\eth}$, which are generalizations of the Laplace
operator on the sphere $L^2$:
\begin{equation}
L^2 f := \frac{1}{\sin \theta} \: \partial_\theta \left(
\sin \theta \: \partial_\theta f \right) 
+ \frac{1}{\sin^2 \theta} \: \partial_\varphi^2 f \; .
\end{equation}
For a function with zero spin weight we in fact find that \mbox{$L^2
f=\bar{\eth} \eth f = \eth \bar{\eth} f$}.

One can also find generalizations of the standard angular momentum
operators for the case of non-zero spin weight by looking for
operators $\hat{J}_z$ and $\hat{J}_\pm$ such that (here we are
ignoring the factor $-i \hbar$ that normally appears in quantum
mechanics)~\cite{Dray85}
\begin{eqnarray}
\hat{J}_z \: {}_sY^{l,m} &=& i\,m \: {}_sY^{l,m} ,
\label{eq:spinspherical_angular1} \\
\hat{J}_\pm \: {}_sY^{l,m} &=& i\, \left[ (l \mp m)(l+1 \pm m)\right]^{1/2}
{}_sY^{l,m \pm 1} . \hspace{8mm} 
\label{eq:spinspherical_angular2}
\end{eqnarray}
One then finds that such operators must have the form
\begin{eqnarray}
\hat{J}_z &=& \partial_\varphi \; , \\
\hat{J}_\pm &=& e^{\pm i \varphi} \Big[ \pm i \partial_\theta - \cot \theta \:
\partial_\varphi - i\, s\, \csc \theta \Big] \; .
\end{eqnarray}

The operators for the $x$ and $y$ components of the angular momentum
are then simply obtained from \mbox{$\hat{J}_\pm = \hat{J}_x \pm\,i
\hat{J}_y$}, so that we find:
\begin{equation}
\hat{J}_x = \left( \hat{J}_+ + \hat{J}_- \right)/2 \; , \quad
\hat{J}_y = - i \left( \hat{J}_+ - \hat{J}_- \right)/2 \; ,
\end{equation}

There are several important properties of the spin-weighted
spherical harmonics that can be obtained directly from their
definition.  In the first place, one can show that the complex
conjugate of ${}_sY^{l,m}$ is given by
\begin{equation}
{}_s\bar{Y}^{l,m} (\theta,\varphi) =
(-1)^{s+m} \: {}_{-s}Y^{l,-m} (\theta,\varphi) \; .
\label{eq:spinspherical_conjugate}
\end{equation}
Also, the different ${}_sY^{l,m}$ are orthonormal:
\begin{equation}
\oint {}_sY^{l,m} (\theta,\varphi) \: {}_{s'}\bar{Y}^{l',m'} (\theta,\varphi)
\: d\Omega = \delta_{ss'} \delta_{ll'} \delta_{mm'} \; ,
\label{eq:spinspherical_orthonormal}
\end{equation}

The integral of three spin-weighted spherical harmonics is also
frequently needed, for example in the calculation of the momentum flux
of gravitational waves, and can be expressed in general as
\begin{eqnarray}
\oint {}_{s_1}Y^{l_1,m_1} (\theta,\varphi) \: {}_{s_2}Y^{l_2,m_2}
(\theta,\varphi) \: {}_{s_3}Y^{l_3,m_3} (\theta,\varphi)\: d\Omega =
\nonumber \\
\left[ \frac{\left( 2l_1+1 \right) \left( 2l_2+1 \right)
\left( 2l_3+1 \right)}{4 \pi} \right]^{1/2} \hspace{20mm} \nonumber \\
\hspace{5mm} \times
\left( \begin{array}{ccc}
l_1  & l_2  & l_3 \\
-s_1 & -s_2 & -s_3
\end{array} \right)
\left( \begin{array}{ccc}
l_1 & l_2 & l_3 \\
m_1 & m_2 & m_3
\end{array} \right) . \hspace{15mm}
\label{eq:spinspherical_three}
\end{eqnarray}
The above expression involves the Wigner 3-$lm$ symbols, which are
related to the standard Clebsch--Gordan coefficients $\left<
l_1,m_1,l_2,m_2 | j_3,m_3 \right>$ through
\begin{eqnarray}
\left( \begin{array}{ccc}
l_1 & l_2 & l_3 \\
m_1 & m_2 & m_3
\end{array} \right) = \hspace{20mm} \nonumber \\
\frac{\left( -1 \right)^{l_1 - l_2 - m_3}}{\sqrt{2\,l_3 + 1}}
\left< l_1,m_1,l_2,m_2 | l_3,-m_3 \right> \; .
\end{eqnarray}

The Clebsch--Gordan coefficients arise from the addition of angular
momentum in quantum mechanics, and correspond to the coefficients of
the expansion of an eigenstate $\left| L,M \right>$ with total angular
momentum $L$ and projection $M$, in terms of a basis formed by the
product of the individual eigenstates $\left| l_1,m_1 \right> \left|
l_2,m_2 \right>$. These coefficients have some important symmetries,
though these symmetries are easier to express in terms of the
\mbox{3-$lm$} symbols. In particular, the \mbox{3-$lm$} symbols are
invariant under an even permutation of columns, and pick up a factor
$(-1)^{l_1+l_2+l_3}$ under an odd permutation.  Also,
changing the sign of all three $m$'s again introduces a factor
of $(-1)^{l_1+l_2+l_3}$

A closed expression for the Clebsch--Gordan coefficients was first
found by Wigner~(see {{\em e.g.}~\cite{Wigner59}). This expression is
somewhat simpler when written in terms of the 3-$lm$ symbols and
has the form

\begin{widetext}
\begin{eqnarray}
&& \left( \begin{array}{ccc}
l_1 & l_2 & l_3 \\
m_1 & m_2 & m_3
\end{array} \right) = (-1)^{l_1-m_1} \:
\delta_{m_1+m_2,-m_3} \nonumber\\ && \hspace{10mm} \times \left[
\frac{(l_1+l_2-l_3)!\:(l_1+l_3-l_2)!\:
(l_2+l_3-l_1)!\:(l_3+m_3)!\:(l_3-m_3)!}
{(l_1+l_2+l_3+1)!\:(l_1+m_1)!\:(l_1-m_1)!\:(l_2+m_2)!\:(l_2-m_2)!}
\right]^{1/2} \nonumber \\ && \hspace{10mm} \times \sum_{k \geq 0}
\frac{(-1)^k}{k!} \left[ \frac{(l_2+l_3+m_1-k)!\:(l_1-m_1+k)!}
{(l_3-l_1+l_2-k)!\:(l_3-m_3-k)!\:(l_1-l_2+m_3+k)!} \right] .
\hspace{10mm}
\label{eq:spinspherical_ClebschJordan}
\end{eqnarray}
\end{widetext}

In the above expression the sum runs over all values of $k$ for which
the arguments inside the factorials are non-negative. Also, if the
particular combination of $\{l_i,m_i\}$ is such that the arguments of
the factorials outside of the sum are negative, then the corresponding
coefficient vanishes. A more symmetric (though longer) expression
that is equivalent to~\eqref{eq:spinspherical_ClebschJordan} was later
derived by Racah~\cite{Racah42}, but we will not write it here.

In the general case equation~\eqref{eq:spinspherical_ClebschJordan} is
rather complicated, but this is not a serious problem as one can find
tables of the most common coefficients in the literature, and even
web-based ``Clebsch--Gordan calculators''.  Moreover, in some special
cases the coefficients simplify considerably.  For example, in the
case when $m_1=l_1$, $m_2=l_2$ and $l_3=m_3=l_1+l_2$ one finds
\begin{equation}
\left( \begin{array}{ccc}
l_1 & l_2 & l_1+l_2 \\
l_1 & l_2 & l_1+l_2
\end{array} \right) = \frac{1}{\sqrt{2(l_1+l_2)+1}} \; .
\end{equation}
Another particularly interesting case corresponds to taking
$l_3=m_3=0$ ({\em i.e.} zero total angular momentum in quantum
mechanics).  In that case we find
\begin{eqnarray}
\left( \begin{array}{ccc}
l_1 & l_2 & 0 \\
m_1 & m_2 & 0
\end{array} \right) &=& \left< l_1,m_1,l_2,m_2 | 0,0 \right>
\nonumber \\
&=& \frac{(-1)^{l_1 - m_1}}{\sqrt{2 l_1 + 1}} \;
\delta_{l_1,l_2} \: \delta_{m_1,-m_2} \; . \hspace{5mm}
\end{eqnarray}

The cases with $l_3=1$ are also interesting as they appear in the
expression for the linear momentum carried by gravitational waves. One
finds, in particular

\begin{widetext}
\begin{eqnarray}
\left( \begin{array}{ccc}
l_1 & l_2 & 1 \\
m_1 & m_2 & 0
\end{array} \right)
&=& (-1)^{l_1-m_1} \: \delta_{m_1+m_2,0}
\left[ \frac{2\,m_1\,\delta_{l_1,l_2} }
{\sqrt{(2l_1+2)\,(2l_1+1)\,(2l_1)}} \right. \nonumber \\
&+& \left. \delta_{l_1,l_2+1}\, \sqrt{ \frac{(l_1+m_1)\,(l_1-m_1)}
{l_1\,(2l_1+1)\,(2l_1-1)} }
- \delta_{l_1+1,l_2} \,\sqrt{\frac{(l_2-m_2)\,(l_2+m_2)}
{l_2\,(2l_2+1)\,(2l_2-1)} }\, \right]\; , \hspace{5mm} \\
\left( \begin{array}{ccc}
l_1 & l_2 & 1 \\
m_1 & m_2 & \pm 1
\end{array} \right)
&=& (-1)^{l_1-m_1} \: \delta_{m_1+m_2,\mp1}\,
\Bigg[ \pm \: \delta_{l_1,l_2}\,\sqrt{ \frac{(l_1 \mp m_1)\,(l_1 \mp m_2)}
{l_1\,(2l_1+2)\,(2l_1+1)} }  \nonumber \\
&+& \delta_{l_1,l_2+1}\, \sqrt{ \frac{(l_1 \mp m_1)\,(l_1 \pm m_2)}
{2l_1\,(2l_1+1)(2l_1-1)}}
+ \delta_{l_1+1,l_2}\, \sqrt{ \frac{(l_2 \mp m_2)\,(l_2 \pm m_1)}
{2l_2\,(2l_2+1)\,(2l_2-1)}}\, \Bigg] \; . 
\end{eqnarray}

\end{widetext}


\bibliographystyle{bibtex/prsty}
\bibliography{bibtex/referencias}


\end{document}